\documentclass[
  ,draft            
  ]
  {aipproc}

\layoutstyle{6x9}

\newcommand\sun{\odot}%

\begin{document}

\title{Quenching Star Formation: Can AGN Do the Trick?}


\author{Jared M. Gabor}{
  address={University of Arizona}
}

\author{Romeel Dav\'{e}}{
  address={University of Arizona}
}

\begin{abstract}
We post-process galaxy star formation histories in cosmological
hydrodynamics simulations to test quenching mechanisms associated with
AGN.  By comparing simulation results to color-magnitude diagrams and
luminosity functions of SDSS galaxies, we examine whether ``quasar
mode'' or ``radio mode'' AGN feedback can yield a realistic red
sequence.  Both cases yield red sequences distinct from the blue
cloud, decent matches to the luminosity function, and galaxies that
are too blue by about 0.1 magnitudes in $g-r$.  Our merger-based
prescription for quasar mode feedback, however, yields a red sequence
build-up inconsistent with observations: the luminosity function lacks
a characteristic knee, and the brightest galaxies include a small
number of young stars.

\end{abstract}

\maketitle


\section{Introduction}

Recent research has linked active supermassive black holes to the
evolution of galaxies and clusters, both observationally via the
$M_{\mathrm{BH}}$--$ \sigma_{\mathrm{bulge}}$ relation, and
theoretically to solve the cooling flow and star formation quenching
problems.  Whether AGN can inject enough energy into the right gas at
the right place and time on galactic scales remains a contentious
question.  But when AGN are invoked in theoretical studies, do they
have the desired \emph{effects} in detail?  If we assume that the
physics of feedback works on small scales, do we reproduce the correct
distributions of galaxy and inter-galactic medium properties?

In this work, we approach this question in terms of galaxy properties.
We test how AGN feedback impacts two observations: a) the bimodality
of galaxy colors, with a tight red sequence, a diffuse blue cloud, and
a discernible gap in between; and b) the galaxy luminosity function
with a characteristic knee at $r \simeq -20.77$, with red galaxies
dominating the bright population and blue galaxies dominating the
faint one \citep{blanton05_lfs}.  The color bimodality requires star
formation to cease in less than 1 Gyr as a galaxy moves from blue to
red.  The knee in the luminosity function specifies a characteristic
luminosity or mass scale above which galaxies tend to stop forming
stars.  AGN feedback in different contexts could drive these phenomena
by quenching star formation.

Two independent ``modes'' of AGN feedback have emerged as potential
star formation quenching mechanisms.  High luminosity AGN or quasars
resulting from galaxy mergers may be able to expel the cold gas from a
galaxy and prevent future star formation (``quasar mode feedback'').
On the other hand, low luminosity radio AGN embedded in massive
galaxies may inject energy into their hot X-ray gas halos, preventing
the gas from cooling and forming stars \citep{croton06}.

We test these mechanisms of star formation quenching using
post-processing techniques on cosmological hydrodynamics simulations.

\section{Analysis: Simulations and Post-processing techniques}
We develop and apply a new set of post-processing routines to existing
cosmological simulations.  We have simulated a $\Lambda$CDM volume of
$(48 h^{-1})^3$ Mpc$^3$ with $256^3$ dark matter $+$ $256^3 $ gas
particles using a modified version of GADGET-2, a smoothed-particle
hydrodynamics code \citep{springel05}.  This choice of volume gives a
gas particle mass of $\sim 10^8 M_{\sun}$, and a galaxy mass
resolution of $\sim 7\times10^9 M_{\sun}$.  Our version of GADGET-2
dynamically incorporates analytic sub-resolution models for star
formation, feedback from star-formation driven winds, and chemical
enrichment \citep{springel03, opp08}.  Simulation outputs consist of a snapshot of the simulated
volume at each of 100 redshifts from $z=30$ to $z=0$.

With our new post-processing routines, we mimic the effects of
different quenching mechanisms by applying simple prescriptions to the
resulting galaxy star formation histories.  As an example, consider
quenching due to merger-induced quasars.  We simplify this to a
quenching criterion: any (resolved) galaxies that are the remnants of
major (3:1 mass ratio or smaller) mergers should not form any new
stars.  We identify major mergers in the simulation outputs as those
galaxies whose stellar mass has grown by at least a factor of
$(1+1/3)$ from one time step to the next ($\sim 300$ Myr for $z< 1$).
A merger remnant remains flagged as such for all later time steps
unless it is subsumed by another more massive galaxy.  We then step
through each time step starting at $z\sim10$ and determine whether
each star particle that formed within the last time step formed within
a merger remnant.  If so, then we flag that star particle as quenched:
it should never have formed.  Then, when we compute stellar masses or
luminosities of galaxies, we just \emph{ignore} those star particles
that we flagged as quenched.

For radio mode quenching we follow an analogous
prescription.  Here, the quenching condition derives from the
characteristic minimum dark matter halo mass required for a stable hot
gas halo, $M_{\mathrm{halo}} \sim 10^{12} M_{\sun}$.  We assume that
no new stars can form within halos above this critical mass because
the AGN prevents the hot gas halo from cooling.

\section{Results}
\begin{figure}
  \includegraphics[height=0.8\textwidth]{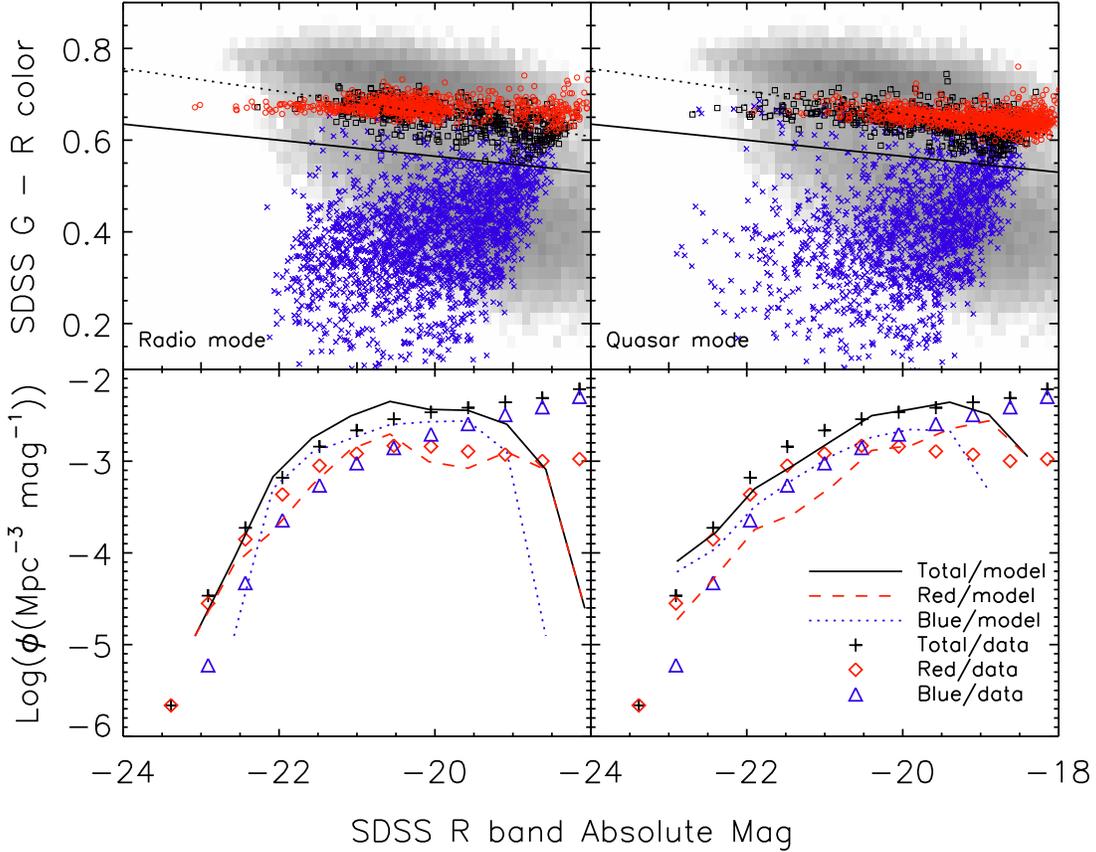}
  \caption{Color-magnitude diagrams and luminosity functions for our
  simulations with quenching prescriptions applied.  Top panels
  compare our simulated CMDs (symbols) to the low-redshift CMD of the
  SDSS VAGC (gray shading).  Galaxy symbols are coded by the age of
  the youngest star particle: 0--1 Gyr (blue x's), 1--4 Gyr (black
  squares), and $>4$ Gyr (red circles).  We include lines separating blue
  and red galaxy loci (solid lines for simulation CMDs, and dotted
  lines for real galaxies).  Bottom panels split luminosity functions
  into blue galaxies (dotted line/triangles), red galaxies (dashed
  line/diamonds), and all galaxies (solid line/plusses) for both
  simulated (lines) and real (symbols) galaxies.  Our mass resolution
  limit, manifest as a diagonal envelope on the right side of the
  CMDs, corresponds roughly to $r=-20$ in the LFs.} \label{fig.cmdlf}
\end{figure}

In Figure \ref{fig.cmdlf} we compare color-magnitude diagrams (CMDs)
and luminosity functions (LFs) of our simulation quenching models to
data from the Sloan Digital Sky Survey (SDSS) Value-Added Galaxy
Catalog (VAGC) low-redshift sample \citep{blanton05_vagc}.  The left
column results from our radio mode quenching mechanism, which is keyed
to the critical dark matter halo mass, $M_{\mathrm{halo}}=10^{12}
M_{\sun}$, while the right column results from our quasar mode
mechanism based on galaxy mergers.

Both star formation quenching mechanisms successfully produce a
bimodality in galaxy colors, whereas simulations without quenching
produce only star-forming blue galaxies (including ones brighter than
$r\simeq -24$; not shown).  The separation between blue and red
galaxies for the simulations (solid line), however, is bluer than that
for real galaxies (dotted line) by about 0.1 magnitudes in $g-r$.
This problem most likely arises because our simulated galaxies are
more metal poor than real galaxies, which may reflect uncertainties in
yields or stellar population synthesis models.

We do not include the effects of dust, which we will explore in
greater detail in the future.  Dust extinction moves simulated blue
galaxies up and to the right in the CMDs, perhaps contaminating the
red sequence with a small population of dusty galaxies.

\subsubsection{Quasar mode and mergers: the wrong red sequence}
The red sequence driven by galaxy mergers fails to reproduce key
characteristics of the observed red sequence in subtle ways.  While
the overall shape of the luminosity function matches reasonably well,
this quenching mechanism produces slight excesses of galaxies at both
the faint and bright ends.  Furthermore, merger quenching yields an
underabundance of red galaxies around $M^*$, leading to a flatter
bright-end slope.  Basically, the spread in merger masses does not
yield a sharp knee at $M^*$.  The growth of the red sequence suggests
additional conflicts with observation: the brightest galaxies include
a population of young stars, acquired via recent merger events with
younger galaxies.  These trace young stellar populations do not appear
in observed bright red sequence galaxies \citep{sanchez09}.

\subsubsection{Radio mode and hot halos: the right red sequence?}
The red sequence driven by hot gas halos that never cool to form stars
shows better overall agreement with the observed luminosity function.
Further, galaxies \emph{at all epochs} tend to migrate to the red
sequence at a rough characteristic absolute magnitude ($r\sim-21$)
directly related to the critical halo mass, $10^{12} M_{\sun}$.
Although variations in the assembly histories of galaxies in halos of
a given mass smear out this characteristic crossing magnitude, the
critical halo mass directly gives rise to the knee in the luminosity
function.  The $M^*$ population of red galaxies continues to grow at
$z=0$ as more galaxies end up in massive halos, while the brightest
red galaxies have grown mostly via minor dry mergers over billions of
years.


\begin{theacknowledgments}
JMG thanks the conference organizers and attendees.  We thank Kristian
Finlator and Ben D. Oppenheimer for guidance and help with our analysis.
\end{theacknowledgments}



\bibliographystyle{aipproc}   

\bibliography{paper}

\begin{thebibliography}{7}
\expandafter\ifx\csname natexlab\endcsname\relax\def\natexlab#1{#1}\fi
\providecommand{\enquote}[1]{``#1''}
\expandafter\ifx\csname url\endcsname\relax
  \def\url#1{\texttt{#1}}\fi
\expandafter\ifx\csname urlprefix\endcsname\relax\def\urlprefix{URL }\fi
\providecommand{\eprint}[2][]{\url{#2}}

\bibitem[{Blanton} et~al.(2005{\natexlab{a}})]{blanton05_lfs}
M.~R. {Blanton}, R.~H. {Lupton}, D.~J. {Schlegel}, M.~A. {Strauss},
  J.~{Brinkmann}, M.~{Fukugita}, and J.~{Loveday}, \emph{\apj} \textbf{631},
  208--230 (2005{\natexlab{a}}), \eprint{arXiv:astro-ph/0410164}.

\bibitem[{Croton} et~al.(2006)]{croton06}
D.~J. {Croton}, V.~{Springel}, S.~D.~M. {White}, G.~{De Lucia}, C.~S. {Frenk},
  L.~{Gao}, A.~{Jenkins}, G.~{Kauffmann}, J.~F. {Navarro}, and N.~{Yoshida},
  \emph{\mnras} \textbf{365}, 11--28 (2006), \eprint{arXiv:astro-ph/0508046}.

\bibitem[{Springel}(2005)]{springel05}
V.~{Springel}, \emph{\mnras} \textbf{364}, 1105--1134 (2005),
  \eprint{arXiv:astro-ph/0505010}.

\bibitem[{Springel} and {Hernquist}(2003)]{springel03}
V.~{Springel}, and L.~{Hernquist}, \emph{\mnras} \textbf{339}, 289--311 (2003),
  \eprint{arXiv:astro-ph/0206393}.

\bibitem[{Oppenheimer} and {Dav{\'e}}(2008)]{opp08}
B.~D. {Oppenheimer}, and R.~{Dav{\'e}}, \emph{\mnras} \textbf{387}, 577--600
  (2008), \eprint{0712.1827}.

\bibitem[{Blanton} et~al.(2005{\natexlab{b}})]{blanton05_vagc}
M.~R. {Blanton}, D.~J. {Schlegel}, M.~A. {Strauss}, J.~{Brinkmann},
  D.~{Finkbeiner}, M.~{Fukugita}, J.~E. {Gunn}, D.~W. {Hogg}, {\v
  Z}.~{Ivezi{\'c}}, G.~R. {Knapp}, R.~H. {Lupton}, J.~A. {Munn}, D.~P.
  {Schneider}, M.~{Tegmark}, and I.~{Zehavi}, \emph{\aj} \textbf{129},
  2562--2578 (2005{\natexlab{b}}), \eprint{arXiv:astro-ph/0410166}.

\bibitem[{Sanchez-Blazquez} et~al.(2009)]{sanchez09}
P.~{Sanchez-Blazquez}, B.~K. {Gibson}, D.~{Kawata}, N.~{Cardiel}, and
  M.~{Balcells}, \emph{ArXiv e-prints}  (2009), \eprint{0908.2548}.

\end{thebibliography}




%
%
%
%
%
\end{document}